\documentclass[traditabstract,letter]{aa} 
\usepackage{graphicx}
\usepackage{txfonts}
\begin{document}
   \title{The $\rm M_{BH}-M_{star}$ relation of
   	obscured AGNs at high redshift
 	\thanks{Based on data obtained at the VLT through the ESO programs
                         73.A-0598, 076.A-0681, 077.B-0368.
	}
	}


   \author{J.E. Sarria\inst{1}
          \and
          R. Maiolino\inst{2}
		  \and
		  F. La Franca\inst{1}
		  \and
		  F. Pozzi\inst{3,4}
		  \and
		  F. Fiore\inst{2}
		  \and
		  A. Marconi\inst{5}
		  \and
		  C. Vignali\inst{3,4}
		  \and
		  A. Comastri\inst{4}
          }

   \institute{Dipartimento di Fisica, Universit\`a Roma Tre,
   			via della Vasca Navale 84, 00146 Roma, Italy
			\and
             INAF - Osservatorio Astronomico di Roma,
			 	via di Frascati 33, 00040 Monte Porzio Catone, Italy
         \and
	Dipartimento di Astronomia, Universit\`a 
		degli Studi di Bologna, Via Ranzani1 , I–40127 Bologna, Italy
         \and
	INAF - Osservatorio Astronomico di Bologna,
	via Ranzani 1, 40127 Bologna, Italy
         \and
		 Dipartimento Fisica e Astronomia,
		 Universit\`a degli Studi di Firenze, Largo E. Fermi 2,
		 50125 Firenze, Italy	
             }

   \date{Received ; accepted }

 
  \abstract
   {
   We report the detection of broad H$\alpha$ emission
   in three X-ray selected obscured AGNs at z$\sim$1--2.
   By exploiting the H$\alpha$ width and the intrinsic X-ray
   luminosity, we estimate their black hole masses, which are in the range
   0.1-3$\rm \times 10^9~M_{\odot}$. By means of multi-band photometric data,
   we measure the stellar mass of their host galaxy and, therefore, infer their $\rm
   M_{BH}/M_{star}$ ratio. These are the first {\it obscured} AGNs at high-z,
selected based on their black hole accretion (i.e. on the basis of their X-ray
luminosity), that can be located on the $\rm M_{BH}$-$\rm M_{star}$ relation at high-z.
All of these obscured high-z AGNs are fully consistent with
the local $\rm M_{BH}$-$\rm M_{star}$
relation. This result conflicts
with those for other samples of AGNs in the same redshift range,
whose $\rm M_{BH}/M_{star}$
ratio departs significantly from the value observed in local galaxies.
We suggest that the
obscured AGNs in our sample are in an advanced evolutionary stage, have already
settled onto the local $\rm M_{BH}$-$\rm M_{star}$ relation, and whose nuclear
activity has been temporarily revived by recent galaxy interactions.
}

   \maketitle
%

   \begin{figure*}[t!]
   \centering
   \includegraphics[width=4.5cm]{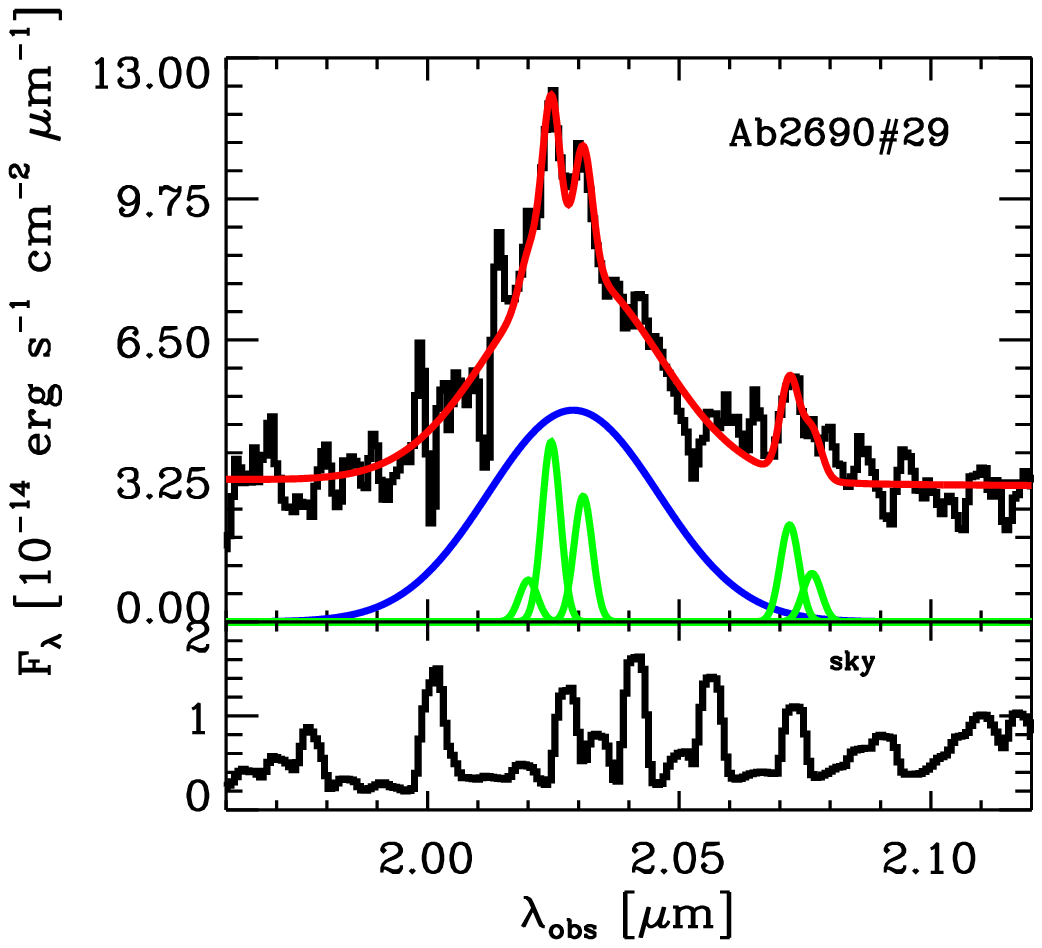}
   \includegraphics[width=4.5cm]{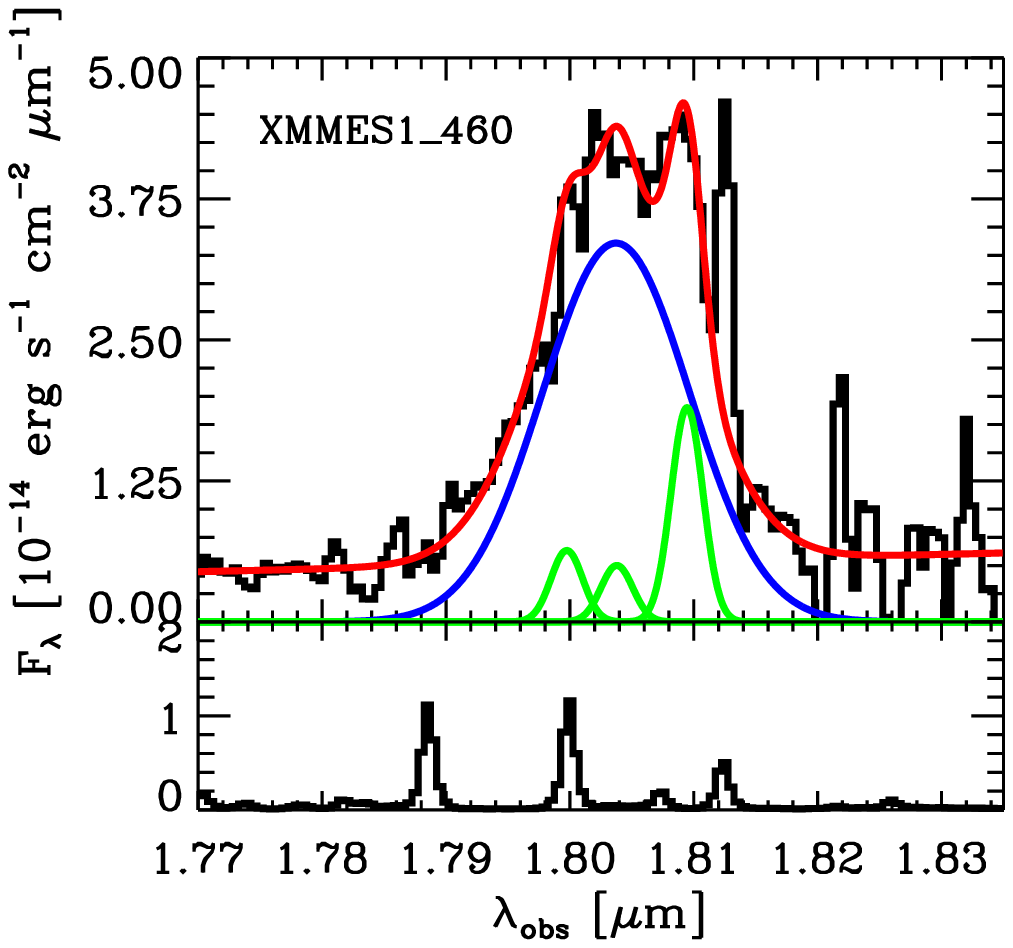}
   \includegraphics[width=4.5cm]{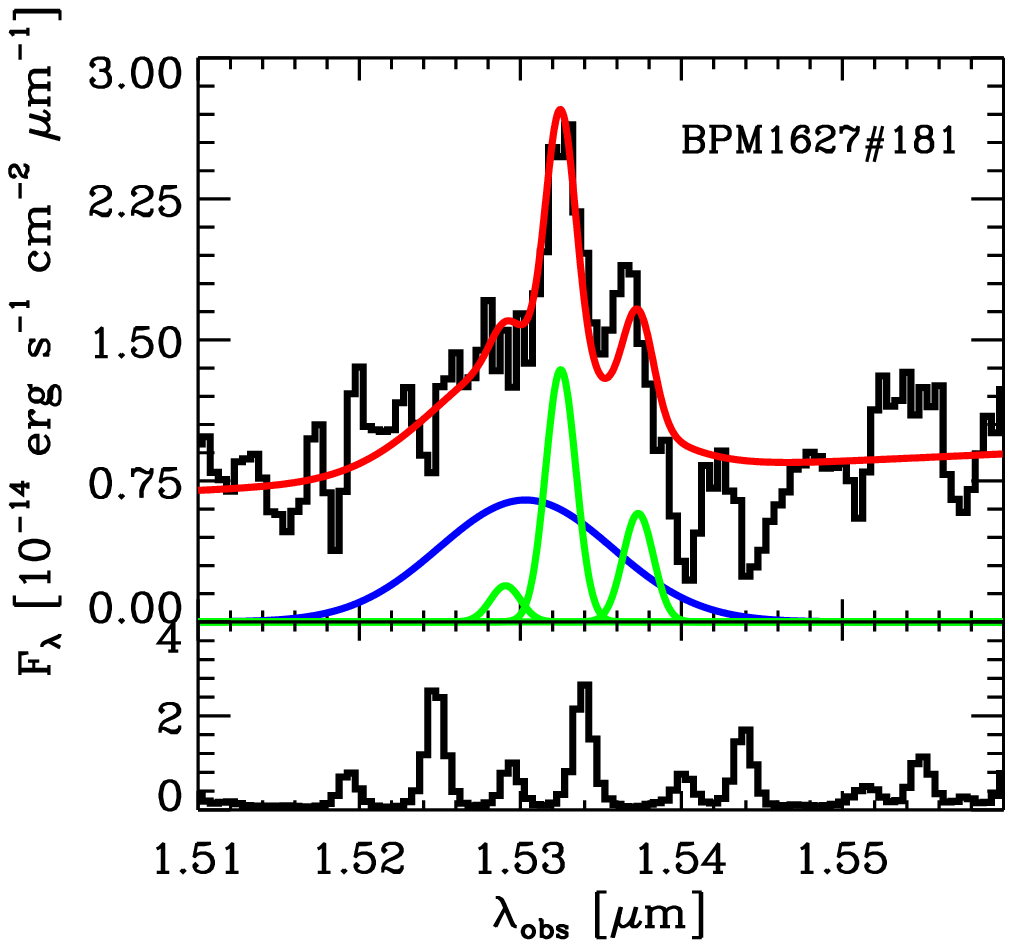}
      \caption{Near-IR spectra (zoomed around H$\alpha$) of the three
	  X-ray selected obscured AGNs with evidence for a broad component
	  of H$\alpha$. The red line shows the fit resulting from a combination
	  of a broad H$\alpha$ Gaussian component (blue line), narrow
	  Gaussian components of H$\alpha$, [NII] and, in the case of Abell2690\#29,
	  [SII] (green lines) and a linear continuum. The bottom panels show the sky spectrum
	  in arbitrary units.
              }
         \label{spectra}
   \end{figure*}

\begin{table*}[!]
\caption{\label{t1} Properties of the X-ray selected obscured AGN with
broad H$\alpha$.}
\centering
\begin{tabular}{lccccccccc}
\hline\hline
 ID	 		& z	 & instr.	& $\rm log L_{2-10keV}$$^a$& $\rm log N_H$ &
 $\rm log L_{bol}$$^b$ & $\rm FWHM_{H\alpha}$$^c$& $\rm log M_{BH}$&
 $\rm L/L_{Edd}$& $\rm log M_{star}$ \\
			& 	&	& $\rm [erg~s^{-1}]$   &$\rm [cm^{-2}]$& $\rm [erg~s^{-1}]$& $\rm [km~s^{-1}]$   &$\rm [M_{\odot}]$&				& $\rm [M_{\odot}]$ \\ 
\hline
Ab2690\#29  & 2.087 & ISAAC & 44.95  			   &	 22.80 	   & 46.81
& 5871$\pm$245
& 9.44$\pm$0.31    & 0.12			& 12.50$\pm$0.20  \\  
XMMES1\_460		& 1.748 & SINFONI & 44.86  			   &	 22.50	   & 46.59
& 2316$\pm$113
& 8.54$\pm$0.31	   & 0.63			& 11.53$\pm$0.20  \\
BPM1627\#181& 1.335 &  SINFONI & 44.20  			   &	 22.81	   & 45.76
& 2491$\pm$511
& 8.15$\pm$0.36	   & 0.32			& 11.28$\pm$0.20  \\
\hline
\end{tabular}
\tablefoot{
\tablefoottext{a}{Absorption corrected 2--10~keV luminosity.}
\tablefoottext{b}{Bolometric luminosity inferred by using the X-ray bolometric
corrections given in Marconi et al. (2004).}
\tablefoottext{c}{Full Width at Half Maximum of the broad component of H$\alpha$
in km/s.}
}
\end{table*}

\section{Introduction}
A major breakthrough in our understanding of galaxy
evolution has been 
the discovery of a tight correlation, in the local universe,
between the mass
of supermassive black holes ($\rm M_{BH}$) and the
mass of their host spheroids
(e.g. Magorrian et al. 1998, Ferrarese \& Merritt 2000).
The existence of this relation implies
a strong physical connection between galaxy formation
and growth of black holes at their centers.
Various models and simulations
have been proposed to explain this correlation (e.g., Menci et al. 2006,
Marulli et al. 2008, Hopkins et al. 2006,
Volonteri \& Natarajan 2009).
These models predict different
evolutionary patterns on the $\rm M_{BH}-M_{star}$
diagram through the cosmic epochs.
Therefore, determining the location of high-z galaxies on
the $\rm M_{BH}-M_{star}$ diagram provides a crucial test
for galaxy-black hole coevolutionary scenarios.

Various observational studies have attempted to measure the evolution of the
$\rm M_{BH}-M_{star}$ relation at high redshift. Most of these studies
infer the $\rm M_{BH}$ in high-z AGNs by
using ``virial estimators'', involving the AGN luminosity and width of
the ``broad emission lines''
(e.g. Vestergaard et al. 2009, Shen et al. 2010).
Broad lines, especially in the UV (observed in the optical at high-z),
are generally detected only in unobscured, type 1 AGNs (AGN1s), whose
broad line region can be observed directly. Thus the
investigation of the $\rm M_{BH}-M_{star}$ relation
has focused mostly on AGN1s, especially at high-z. Jahnke et al. (2009)
use the virial relations
to measure the BH masses of ten type 1 AGN at z$\sim$1.4 and derive
the stellar masses of their host galaxy by means
of multi band fitting. They find
that the BH-to-total stellar mass ratio does not evolve relative to the local
relation; however, since several hosts show evidence of a substantial disk
component they suggest that the ratio of BH mass to bulge mass (given by the
difference between total and disk stellar mass) probably
evolves with redshift.
Merloni et al. (2010) expand the Jahnke et al. sample to
89 X-ray selected type 1 AGN at 1$<$z$<$2.4 in COSMOS (adopting a different method
to measure the stellar masses). They infer
that the $\rm M_{BH}/M_{star}$ ratio {\it increases} significantly,
by a factor of about two, relative to the local relation. A similar result
was obtained by Peng et al. (2006) by using a sample of lensed quasars
in a similar redshift range. At higher redshifts (z$\sim$4--6), dynamical mass
measurements, obtained by exploiting CO maps, suggest
that the $\rm M_{BH}/M_{star}$ ratio
in type 1 AGNs increases even further, up to
an order of magnitude relative to the local relation
(Walter et al. 2004; Maiolino et al. 2007a, Lamastra
et al. 2010, Wang et al. 2010).

A possible caveat of these observational studies
is that the targets are extracted from AGN-selected samples,
i.e. on the basis of the black-hole accretion
rate.
This is expected to introduce a bias in favor of massive black holes
(which can reach higher absolute accretion rates within their Eddington limit)
\footnote{Quasars with dynamical masses inferred from CO maps
are also selected based on their millimeter continuum, hence for this subsample
the bias may be more complex.}.
More specifically, even if the $\rm M_{BH}-M_{star}$ relation does not
evolve with redshift, it does have a scatter and the AGN-selection bias
favors the selection of objects with $\rm M_{BH}/M_{star}$ higher
than the true distribution, hence mimicking an evolution.
Lauer et al. (2007) estimate that this bias may
increase the inferred $\rm M_{BH}-M_{star}$ even by a factor of a few, depending on the
AGN luminosity and 
the intrinsic scatter of the $\rm M_{BH}-M_{star}$ relation at high
redshift. However, Merloni et al. (2010) demonstrate that, at least for
their X-ray selected,
unobscured AGN 1 sample, this bias should not significantly affect
their own results.

The only work investigating the evolution of the $\rm M_{BH}-M_{star}$ relation
through a sample of objects not pre-selected among AGNs is that of
Alexander et al. (2008). Their parent sample is composed of sub-millimeter
galaxies (SMGs), which are strongly starbursting systems at z$\sim$2.
Their sample consists mainly of obscured AGNs hosted such systems.
For a subsample
of these obscured AGNs, which show some broad H$\alpha$ or H$\beta$,
they are able infer the BH mass.
In contrast to the type 1 AGN studies mentioned above,
the inferred $\rm M_{BH}/M_{star}$ ratio of SMGs at z$\sim$2 is {\it lower}
than observed locally. However, this sample may be subject to an opposite
bias than AGN-selected samples. Given the extraordinarily high star-formation
rate of SMGs and the correlation between SFR and stellar mass,
the sample used by Alexander et al. (2008) may be biased toward
objects of high stellar mass (although SMGs are outliers on
the SFR-mass relation, toward higher SFRs).
The sub-mm selection may also favor dust-rich systems
(Santini et al. 2010),
hence more-evolved host galaxies and more massive galactic hosts.
Altogether, these effects may favor
of objects with $\rm M_{BH}/M_{star}$ ratios lower than the real distribution.

Here we present near-IR spectra of three {\it obscured} quasars at
z$\sim$1--2, extracted from a sample of hard X-ray selected AGNs, that display
a broad component of H$\alpha$, allowing us to infer the BH mass and
$\rm M_{BH}/M_{star}$ ratio.
These are the first {\it obscured} AGNs at high-z
selected on the basis of their black hole accretion (i.e. because X-ray
selected) that can be located on the $\rm M_{BH}-M_{star}$ diagram at high-z.
We illustrate that the trend of these obscured AGNs differs
significantly from that of
the other samples of AGNs investigated in the same redshift range.

\section{Sample selection, observations, and data analysis}

These near-IR spectroscopic observations are part of a program
designed to determine the
redshift of and characterize obscured AGNs at high redshift selected
from wide-area hard X-ray surveys. These AGNs are too faint at optical
wavelengths to be observed spectroscopically. The parent samples consist
of XMM hard X-ray sources from the HELLAS2XMM extended survey
(Cocchia et al. 2007) and the XMM survey of the ELAIS-S1 field
(Feruglio et al. 2008). From these samples, we selected 14 sources
characterized by a X-ray to optical flux ratio X/O$>$10, which is typical of
most obscured (Compton-thin) AGNs at high redshift and,
in particular, obscured QSOs (Fiore et al. 2003). Most of these optically
faint AGNs have also very red optical-to-near-IR colors and, more
specifically, $\rm R-K>5$ (Mignoli et al. 2004), i.e. are extremely red objects
(EROs).
This class of objects is generally found to consist mostly of either
quiescent, evolved
galaxies or dust-reddened star-forming galaxies at z$\sim$1--2, with a fraction
of obscured AGNs (e.g. Campisi et al. 2009). In the
specific case of X-ray selected EROs, Mignoli et al. (2004) find that 
most of them are resolved in the K-band with typical sizes of $0''.5$ and
 elliptical-like profiles, implying that the K-band light
is generally dominated by an early type host.
The latter result is also confirmed in
most objects by a SED analysis (Pozzi et al. 2007, 2010).

Our observations were performed in three observing runs.
In the first run, we used ISAAC at the VLT
to observe three HELLAS2XMM sources.
We used ISAAC in its low spectral resolution mode (R$\sim$500), covering
the bands J, H, and K bands.
These observations were presented in Maiolino et al. (2006) and we
refer to that paper for further details. In the second run, we used
SINFONI, the near-IR integral field spectrometer at the VLT, to observe
three HELLAS2XMM sources.
We used both the H+K R$\sim$1500 grating and the J-band R$\sim$2000
grating. The on-source integration times ranged from 40 min to 4 h per
target
in each band. In the third run, we observed nine sources from the ELIAS-S1 sample
with SINFONI, with the same instrumental setting as for the previous SINFONI
run and similar integration times. The seeing during the observations
ranged from $0''.7$ to $1''.5$.
Data reduction was performed using the ESO pipeline.

All targets were detected in the continuum. However,
only for seven of them could we identify emission lines that allow us to
unambiguously determine their redshift. At the inferred redshift
(1$<$z$<$2.1) the inferred X-ray luminosity and X-ray spectral shape
imply that these are all type 2, obscured QSOs
($\rm L_{2-10keV}>10^{44} erg~s^{-1}$, $\rm N_H>10^{22}cm^{-2}$).
Here we focus on three
sources for which we detect a broad component of H$\alpha$ that
allowed us to infer their BH masses.
Their main observational and
physical properties are reported in Table 1.

   \begin{figure}
   \centering
   \includegraphics[width=9cm]{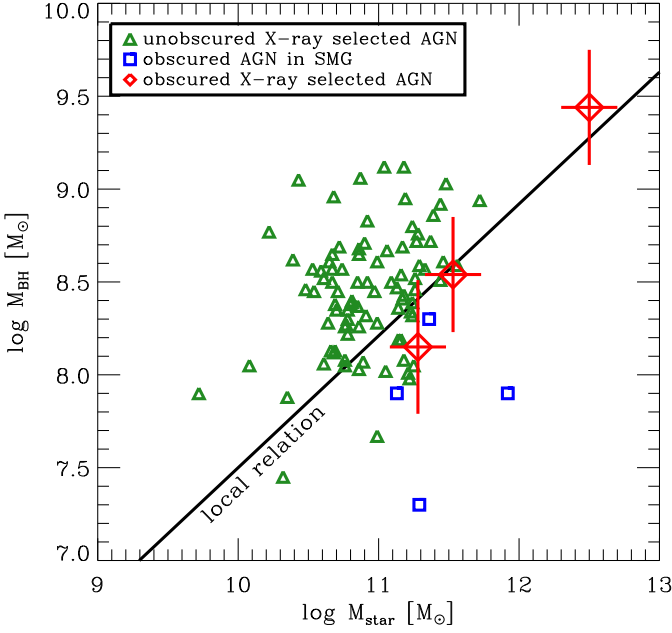}
      \caption{$\rm M_{BH}-M_{star}$ relation for AGNs at z$\sim$1--2.
	  The red diamonds with errorbars show the location of the three X-ray
	  selected obscured AGNs presented in this paper. Green triangles are
	  type 1, X-ray selected unobscured AGNs in COSMOS from Merloni et al. (2010).
	  Blue squares are obscured AGNs in SMGs from Alexander et al. (2008)
	  with broad H$\alpha$ and stellar masses from Hainline et al. (2010).
	  The black line is the local relation obtained by Sani et al. (2010).
              }
         \label{mbh_mst}
   \end{figure}


\section{The $\rm M_{BH}-M_{star}$ relation}

In Fig.~\ref{spectra}, we show the near-IR spectra of the three obscured quasars
at z$\sim$1--2 with broad H$\alpha$ (zoomed around the H$\alpha$ line).
No H$\beta$ broad line is detected for any of these sources, which is
indicative of
a type 1.9 optical classification and consistent with the absorption
inferred from both their X-ray spectrum and their optical faintness.
We fitted the line profile with
a broad Gaussian and three narrow Gaussians (FWHM$<$800~km/s) accounting
for the contribution of the narrow components of both H$\alpha$ and 
[NII]6548,6584. In the case of Abell2690\#29, we also included the [SII]6720
doublet because
the H$\alpha$ is so broad that it extends to this rest-frame wavelength.
We assumed that the narrow components have the same width
and velocity shift. We also assumed that F([NII]6584)/F([NII]6548)$=$3,
as given by the relative transition probabilities of the corresponding levels.
The continuum in this narrow redshift range
is fitted simply with a linear component. Regions affected by very low S/N or
bad residuals due to strong sky lines or bad atmospheric transmission were
excluded from the fit.
The resulting spectral fits are shown in Fig.~\ref{spectra} and the width
of the broad components are listed in Table 1.
The uncertainty in the width of the broad line is also estimated by
accounting for the correlation with the other components.

We cannot use the virial formula for the BH mass based on the H$\alpha$ FWHM and
luminosity proposed by Greene \& Ho (2005) since the H$\alpha$ is likely
absorbed, hence its intrinsic luminosity is difficult to measure.
However, we can use both the virial formula for the BH mass
involving the optical
continuum luminosity $\rm \lambda L_{\lambda}(5100\AA)$ and
$\rm FWHM_{H\beta}$ (Marconi et al. 2010)
\begin{equation}
\rm M_{BH} = 6.16\times 10^6
~\left(\frac{\lambda L_{\lambda}(5100)}{10^{44}~erg~s^{-1}}\right) ^{0.5}
~\left( \frac{FHWM_{H\beta}}{10^3~km~s^{-1}}\right) ^2 ~M_{\odot}
\label{virial_old}
\end{equation}
and the relation between
$\rm L_{2-10keV}$ and $\rm \lambda L_{\lambda}(5100)$ (Eq.5 in Maiolino
et al. 2007b) to estimate the optical continuum, and replace $\rm FWHM_{H\beta}$
with $\rm FWHM_{H\alpha}$ by using the relation given in Greene \& Ho (2005)
(Eq. 3 therein). By replacing these relations in Eq.~\ref{virial_old}, we
obtain
\begin{equation}
\rm M_{BH} = 1.56\times 10^7
~\left(\frac{L_{2-10keV}}{10^{44}~erg~s^{-1}}\right) ^{0.694}
~\left( \frac{FHWM_{H\alpha}}{10^3~km~s^{-1}}\right) ^{2.06} ~M_{\odot} ~.
\label{virial_new}
\end{equation}
Since we know the absorption-corrected 2-10~keV luminosity of our three
sources (Table 1), the width of the broad H$\alpha$ inferred from
our near-IR spectra allows us to infer the BH by exploiting
Eq.~\ref{virial_new}.
The inferred BH masses are reported in Table 1, and range between
$\rm 1.3\times 10^8~M_{\odot}$
and $\rm 2.8\times 10^9~M_{\odot}$. The uncertainty in $\rm M_{BH}$ is
dominated by the dispersion in the $\rm M_{BH}-\sigma,L_{bul}$ local relation (0.3 dex), which is
used to calibrate the virial relations. The uncertainty in the $\rm
L_{2-10keV}-L_{5100}$ relation coefficients also contributes with 0.07 dex (in quadrature), while
the uncertainty in the width of the broad lines contributes significantly only in BPM1627\#181 with
0.19 dex (in quadrature).

For consistency, we checked that the masses inferred by using the H$\alpha$ luminosity
and the relation proposed by Greene \& Ho (2005) imply lower BH masses, confirming
that H$\alpha$ is absorbed. Similarly, the direct use of the relation in Eq.\ref{virial_old}
involving $\rm  L_{\lambda}(5100)$, by taking the observed optical (B rest-frame) continuum
of our sources, deduces BH masses lower than those presented in Table 1,
confirming that the AGN optical light is absorbed and that the observed rest-frame blue
continuum is dominated by the host galaxy (as inferred from the SED fitting).

The BH masses measured by ourselves imply that these three obscured
AGNs are accreting at a rate that is 
about $0.1-0.6$ of their Eddington limit (Table 1).

The stellar masses were inferred by exploiting the multiwavelength
photometric points available for these sources and the
SED fitting code (Fritz et al. 2006) used in Pozzi et al. (2010) and Vignali et
al. (2009) for a subsample
of obscured AGN in the HELLAS2XMM sample.
This code combines synthetic stellar libraries,
AGN dusty torii models (Fritz et al. 2006) including the intrinsic AGN component
(if visible), and starburst IR templates to reproduce the observed SED. We refer to
Pozzi et al. (2010) for a detailed description of the method. Here we only
emphasize that in the case of obscured AGNs, the rest-frame optical to near-IR
radiation is generally dominated by the stellar light from the host galaxy
and therefore constraining the stellar mass is easier than for
the host galaxies of type 1 AGNs. The stellar masses inferred for the host
galaxies of the three obscured quasars presented here are reported in Table 1.
We note that to compare with the local relation, which is between the BH mass and the
stellar mass of the spheroid, we would need to extract the
{\it spheroidal} component of the stellar mass in our high-z targets. However, Mignoli et al.
(2006) demonstrate that
the class of targets in our sample is generally characterized by elliptical-like profiles,
hence the bulk of the stellar light is associated with a massive spheroid.

The resulting location of the three X-ray selected, obscured AGNs at z$\sim$1--2
is shown in Fig.~\ref{mbh_mst} (red diamonds)
along with the location of type 1, unobscured
AGNs in the same redshift range from Merloni et al. (2010) (green triangles)
and the obscured
AGNs in SMGs at z$\sim$2 (blue squares). For the SMG sample we only plot
those AGNs with broad Balmer lines detection
(Alexander et al. 2008) and with available stellar masses from
Hainline et al. (2010).
The solid line shows the local relation deduced by Sani
et al. (2010).
All of the X-ray selected obscured AGNs are fully consistent with the local
$\rm M_{BH}-M_{star}$ relation and do not show any evidence of evolution,
in contrast to both type 1,
unobscured X-ray selected sample and the obscured AGNs
selected among SMGs.
For our obscured AGNs we find that
$\rm \langle \log{(M_{BH}/M_{star})}\rangle = -3.06\pm 19$, which is fully consistent
with the local ratio $\rm \log{(M_{BH}/M_{star})}_{local}=-3.01$
obtained by Sani et al. (2010) in a similar mass range.

\section{Discussion}

Our sample is small and does not allow us to draw firm conclusions
from a statistical point of view. However, it is very intriguing that,
in contrast to other high-z AGN samples, the three X-ray selected
obscured AGNs do not show any systemic deviation from the local
$\rm M_{BH}-M_{star}$ relation.
If this trend were confirmed for a larger sample of X-ray selected obscured
AGNs, it would have important implications for understand either biases
affecting current studies or the evolution of the
$\rm M_{BH}/M_{star}$ ratio during different evolutionary stages
of galaxies.

SMGs have a large scatter, but on average they tend to be
below the local $\rm M_{BH}-M_{star}$ relation (Alexander et al. 2008).
The difference with our results may be explained in terms of the bias
discussed in the introduction, i.e. SMG AGN hosts are biased toward higher
stellar masses because of the mass-SFR relationship and/or because of their
higher dust content (hence more evolved hosts). Alternatively, the
two samples may trace different evolutionary stages.
According to many theoretical models, black hole accretion occurs
predominantly in the phases of galaxy merging and interactions (which
destabilize the gas, allowing it to flow towards the nucleus).
Lamastra et al. (2010) suggest that SMGs may be objects
where previous secular star formation has increased the
stellar mass, while the lack of previous mergers and interactions
has prevented the BH mass from growing significantly. This may explain their lower
BH-to-stellar mass ratio. According to this scenario, SMGs are being observed during
their first major-merging event. In contrast,
X-ray selected, obscured
AGNs (at least in our ERO subsample) may be the descendants of SMGs,
observed at a later stage when, at the end of the whole
interaction/merging phase, galaxies
have already settled onto the local $\rm M_{BH}/M_{star}$ relation.
Our targets have in general elliptical-like profiles (Mignoli et
al. 2005) and an SED
typical of quiescent galaxies (Pozzi et al. 2010), therefore supporting the latter scenario.
However, we that a few
objects in the parent sample have strong far-IR
and submm emission indicative of vigorous star-forming activity
(Vignali et al. 2009, Pozzi et al. 2010). In the specific case of our three targets,
the optical to near-IR photometric points do not allow us to provide tight constraints on
the SFR.

The differences between the location on the $\rm M_{BH}-M_{star}$ diagram
of high-z X-ray selected {\it obscured} AGNs and X-ray
selected {\it unobscured} AGNs is more puzzling. Except for our sample being obscured, our
selection criteria do not differ significantly
from the Merloni et al. (2010) sample (e.g. similar
X-ray luminosity range), hence should not produce differential selection
effects (in particular, our sample should also be unaffected
by the Lauer et al. 2007 bias).
However, we note that the distribution of unobscured AGNs has a large
scatter. The two high-z obscured AGNs with low stellar masses (i.e. overlapping with
the Merloni et al. sample in terms of stellar mass) may at least still be consistent with the
tail of the distribution of unobscured AGNs in the same redshift range.
Larger number statistics are required for the high-z obscured sample to clarify whether
they are, on average, offset or consistent with the unobscured samples in terms
of black hole-to-galaxy mass ratio. Were future observations to confirm
the difference between unobscured and obscured (X-ray selected) AGNs,
this may suggest that the two classes are associated with different evolutionary stages.
The general expectation of models is that
unobscured, type 1 AGNs represent a later evolutionary stage than
to the obscured growth phase. However, the finding that unobscured AGN
are offset from the local $\rm M_{BH}-M_{star}$ relation and are moving towards it (Merloni
et al. 2010), while obscured AGN have already settled onto the local relation,
suggests that the latter are in a later evolutionary stage. A possibility
is that the obscured AGNs in our parent sample of EROs are in a late phase
where the BH and
their host galaxy have already reached the local $\rm M_{BH}-M_{star}$ relation, 
but are temporarily rejuvenated by a late accretion episode, making them
detectable as quasars. In support of this scenario,
we mention that our data reveal that
more than 25\% of the targets in our parent sample
(and all of the sources presented here) are in close
interaction with one or more galaxies located within
a radius of $\sim 15$ kpc, which may be responsible for triggering BH accretion.


A larger sample of high-z obscured, X-ray selected AGNs with broad H$\alpha$ (to
infer the BH mass) is certainly required to confirm with higher statistical significance
that this class
of targets follow the local $\rm M_{BH}-M_{star}$ relation and that they differ from
other high-z AGN samples, as well as to test the various possible scenarios discussed above.

\begin{acknowledgements}
	  We thank C. Feruglio for helping with an early analysis of
	  the data. We thank A. Lamastra and D. Alexander for useful
	  comments.
\end{acknowledgements}

\end{document}